\documentclass[12pt]{article}
\usepackage{amsmath}
\usepackage{graphicx}
\usepackage{float}
\begin{document}

\title{{\bf Finite Upper Bound\\ for the Hawking Decay Time\\ of an Arbitrarily Large Black Hole\\ in Anti-de Sitter Spacetime}
\thanks{Alberta-Thy-13-15, arXiv:yymm.nnnnn [hep-th]}}

\author{
Don N. Page
\thanks{Internet address:
profdonpage@gmail.com}
\\
Theoretical Physics Institute\\
Department of Physics\\
4-183 CCIS\\
University of Alberta\\
Edmonton, Alberta T6G 2E1\\
Canada
}

\date{2015 July 9}

\maketitle
\large
\begin{abstract}
\baselineskip 15 pt

In an asymptotically flat spacetime of dimension $d > 3$ and with the Newtonian gravitational constant $G$, a spherical black hole of initial horizon radius $r_h$ and mass $M \sim r_h^{d-3}/G$ has a total decay time to Hawking emission of $t_d \sim r_h^{d-1}/G \sim G^{2/(d-3)}M^{(d-1)/(d-3)}$ which grows without bound as the radius $r_h$ and mass $M$ are taken to infinity.  However, in asymptotically anti-de Sitter spacetime with a length scale $\ell$ and with absorbing boundary conditions at infinity, the total Hawking decay time does not diverge as the mass and radius go to infinity but instead remains bounded by a time of the order of $\ell^{d-1}/G$.

\end{abstract}

\baselineskip 22 pt

\newpage

Hawking radiation \cite{Hawking:1974rv,Hawking:1974sw} causes a black hole to decay away if there is not sufficient incoming radiation to prevent this.  In four-dimensional asymptotically flat spacetime with no incoming radiation, a nonrotating black hole of at least a solar mass $M_\odot$ (which emits essentially only photons and gravitons, assuming the lightest neutrino is not greatly lighter than the next lightest) has a lifetime $1.1589\times 10^{67}(M/M_\odot)^3$ years \cite{Page:1976df,Page:1976ki,Page:2004xp}, which grows as the cube of the mass and hence diverges when the mass it taken to infinity.

A common toy model for preventing black hole decay is to put it into asymptotically anti-de Sitter spacetime (AdS, which for brevity I shall use not just as an abbreviation for the anti-de Sitter spacetime but also for an asymptotically anti-de Sitter spacetime) with a length scale $\ell$ not too large and imposing reflecting or thermal boundary conditions at spatial infinity \cite{Hawking:1982dh,Witten:1998qj,Witten:1998zw}.  Although AdS has an infinite volume, the gravitational potential that rises indefinitely as one goes to large distances acts effectively like a finite confining box for the black hole and its Hawking radiation.  Massless radiation can escape to infinity, but one can impose reflecting boundary conditions there, and then the radiation reflects back inward in a finite time as seen by observers at finite distances from the center or from a black hole.  Alternatively, one can postulate a sufficiently large thermal bath at or beyond the boundary at radial infinity (though there is not space for a large enough bath within the AdS spacetime itself) that emits thermal radiation inward to keep a sufficiently large black hole from evaporating.

Much less attention has been paid to the possibility of imposing absorbing boundary conditions at infinity in AdS.  Here I consider the case in which one starts in AdS with a large black hole (horizon radius $r_h$ much larger than the AdS length scale $\ell$) and has its Hawking radiation being absorbed by the boundary at infinity, so that no radiation comes back to keep the black hole from evaporating.  I shall then calculate the time for the black hole radius to drop from one initial large value to another.  Surprisingly, it turns out that this time does not diverge even when the initial black hole size (and hence mass) is taken to infinity.

This calculation is simplified by the fact that the geometric optics approximation is good for the bulk of the thermal Hawking radiation emitted by a large black hole, $r_h \gg \ell$.  The geometric optics approximation breaks down for small black holes, $r_h < \ell$, so without doing numerical calculations that are left for the future, I am not able to give a precise estimate for the time for the black hole to evaporate all the way down to zero size.  However, the time to evaporate from $r_h = \ell$ to $r_h = 0$ can be estimated to be of the same order of magnitude as the time to evaporate from $r_h = \infty$ to $r_h = \ell$, namely $t \sim \ell^{d-1}/G$, so that the total time to evaporate from infinite size to zero size will be finite, of the order of $t \sim \ell^{d-1}/G$.

I am using units in which $\hbar = c = k_{\rm{Boltzmann}} = 1$ but shall generally not set Newton's constant $G$ equal to one.  In $d$ dimensions, $G$ has the dimensions of length or time or inverse energy to the $d-2$ power.  I am assuming that the AdS has a length scale $\ell$ very large in comparison with the Planck length one would get by setting $G=1$, so that $\ell^{d-2} \gg G$.  Then the Hawking decay time is $\sim \ell^{d-1}/G \gg \ell$, so the decay time, while finite, is much longer than the time $\pi\ell$, measured along any timelike geodesic in pure anti-de Sitter spacetime, for a null geodesic to travel from that timelike geodesic to infinity and be reflected back to the timelike geodesic.

AdS in $d$ spacetime dimensions with length scale $\ell$ and with a single static spherically symmetric black hole (Schwarzschild-AdS) has the metric
\begin{equation}
ds^2 = - V dt^2 + V^{-1} dr^2 + r^2 d\Omega_{d-2}^2,
\label{AdS-BH metric}
\end{equation}
where
\begin{equation}
V = \frac{r^2}{\ell^2} + 1 - \frac{\mu}{r^{d-3}}
  = \frac{r^2}{\ell^2} + 1
      - \left(\frac{r_h^2}{\ell^2} + 1\right)\frac{r_h^{d-3}}{r^{d-3}},
\label{V}
\end{equation}
with $r_h$ being the value of the radial coordinate $r$ (the circumference divided by $2\pi$ for the $(d-2)$-spheres of symmetry) at the event horizon, where $V$ vanishes, giving
\begin{equation}
\mu = \left(\frac{r_h^2}{\ell^2} + 1\right)r_h^{d-3},
\label{mu}
\end{equation}
and where $d\Omega_{(d-2)}^2$ is the metric on a unit round $(d-2)$-sphere, which has $(d-2)$-dimensional area that I shall denote by the same symbol without the $d$ and without the exponent 2, though this is a numerical value rather than a metric,
\begin{equation}
\Omega_{(d-2)} = \frac{2\pi^\frac{d-1}{2}}{\Gamma\left(\frac{d-1}{2}\right)}.
\label{Omega}
\end{equation}

The surface gravity $\kappa$ of the static black hole is the value on the horizon of the derivative of $\sqrt{-g_{00}} = V^{1/2}$ with respect to the proper radial distance $\sqrt{g_{rr}}dr = V^{-1/2}dr$ and hence is $(1/2) dV/dr$ evaluated on the horizon, $r = r_h$.  Then the Hawking temperature $T$ of the black hole is the surface gravity divided by $2\pi$, which for this Schwarzschild-AdS black hole is
\begin{equation}
T = \left(\frac{d-3}{4\pi}\right)\frac{1}{r_h}
   +\left(\frac{d-1}{4\pi}\right)\frac{r_h}{\ell^2}.
\label{T}
\end{equation}

The first term is larger for a `small' black hole, one with $r_h < \sqrt{(d-3)/(d-1)}\ell$, and gives the Hawking temperature for a Schwarzschild black hole in asymptotically flat spacetime, which is the limit of AdS with $\ell \rightarrow \infty$.  Such a `small' black hole has negative specific heat, Hawking temperature $T$ decreasing with increasing $r_h$ (which is monotonically increasing with the mass $M$ that is proportional to $\mu$), and hence is unstable when one imposes thermal boundary conditions at the corresponding value of $T$ at infinity \cite{Hawking:1982dh}.

The second term is larger for a `large' black hole, one with $r_h > \sqrt{(d-3)/(d-1)}\ell$, which has positive specific heat, Hawking temperature $T$ increasing with increasing $r_h$ and with the mass $M$.  It is stable when one imposes thermal boundary conditions at the corresponding value of $T$ at infinity.  Note that at the boundary between a `small' and a `large' black hole, at $r_h = \sqrt{(d-3)/(d-1)}\ell$, the temperature has its minimum value, $T = \sqrt{(d-3)(d-1)}/(2\pi\ell)$.  One can have thermal equilibrium with a black hole only for temperatures greater than this minimum value (and only for the larger black hole solution with this temperature).  For more details in the 4-dimensional case, see \cite{Hawking:1982dh}.

The Bekenstein-Hawking entropy $S$ for the black hole is its horizon surface area $A_{d-2} = \Omega_{(d-2)}r_h^{d-2}$ divided by $4G$, or
\begin{equation}
S = \frac{\Omega_{(d-2)}r_h^{d-2}}{4G}
  = \frac{\pi^{\frac{d-1}{2}}r_h^{d-2}}{2\Gamma\left(\frac{d-1}{2}\right)G}.
\label{S}
\end{equation}
By integrating $dM = TdS$, one can then get that the black hole mass is
\begin{equation}
M = \frac{(d-2)\Omega_{(d-2)}}{16\pi G}
     \left(\frac{r_h^{d-1}}{\ell^2}+r_h^{d-3}\right)
  = \frac{(d-2)\Omega_{(d-2)}\mu}{16\pi G}.
\label{M}
\end{equation}

Now we need to calculate the Hawking emission rate.  Massive particles cannot reach infinity in AdS but fall back into the black hole, so only massless particles contribute to the decay.  We shall make a geometrical optics approximation and then show that it is valid for the Hawking radiation from a very large black hole, $r_h \gg \ell$.  In the geometric optics approximation, massless quanta move along null geodesics.  One can orient the angular coordinates so that a null geodesic has only $t$, $r$, and one angular coordinate, say $\theta$ with $g_{\theta\theta} = r^2$, changing.  If one normalizes the affine parameter $\lambda$ so that $d/d\lambda$ is the momentum $p$ of the quantum, then the conserved energy is $\omega = -p_0$ and the conserved angular momentum is $L = p_\theta$.   Then one gets
\begin{equation}
\left(\frac{dr}{d\lambda}\right)^2 = \omega^2 - L^2\frac{V}{r^2}.
\label{null geodesic}
\end{equation}

If the null geodesic is coming from infinity to approach the black hole, so that $r$ is decreasing, the geodesic will enter the hole if there is no turning point where $(dr/d\lambda)^2$ vanishes.  This will be the case if the square of the impact parameter $b$, that is $b^2 = L^2/\omega^2$, is smaller than the maximum value of $r^2/V$, which is the square of the critical impact parameter, $b_c^2$.  The maximum value of $r^2/V$ occurs at the minimum value of
\begin{equation}
\frac{V}{r^2} = \frac{1}{\ell^2} + \frac{1}{r^2} - \frac{\mu}{r^{d-1}},
\label{v/r^2}
\end{equation}
which is at
\begin{equation}
r = r_c = \left[\frac{(d-1)\mu}{2}\right]^\frac{1}{d-1}
  = r_h \left[\frac{(d-1)}{2}
  \left(\frac{r_h^2}{\ell^2}+1\right)\right]^\frac{1}{d-1}
\label{r_c}
\end{equation}
and gives the critical impact parameter as
\begin{equation}
b_c = \ell\left\{1+\frac{d-3}{d-1}\frac{\ell^2}{r_h^2}
\left[\frac{d-1}{2}\left(\frac{r_h^2}{\ell^2}+1\right)\right]^{-\frac{2}{d-3}}
\right\}^{-\frac{1}{2}}.
\label{b_c}
\end{equation}

A quantum with typical energy of the order of the Hawking temperature, $\omega = T$, which has the maximum angular momentum $L = b_c\omega$ that can fall into the black hole with that energy, will have angular momentum $L = b_c T$.  For $r_h < \ell$, this is of the order of unity, so that there are not many different angular momentum modes that can fall into the black hole freely with energy $\omega \sim T$, and even the ones that can have wavelengths comparable to the size of the sphere at $r = r_c$.  That implies that the geometric optics approximation is not valid for $r_h < \ell$, so to get the precise values for the Hawking emission rate for `small' black holes in AdS (all black holes in asymptotically flat spacetime), one needs to resort to numerical calculations, such as those in \cite{Page:1976df,Page:1976ki,Page:2004xp}.  I am not aware that any analogous calculations have been done for over the full frequency range (cf.\ \cite{Harmark:2007jy}) for small but nonzero values of $r_h/\ell$ in AdS, and I do not have time to do them for this paper.

However, for `large' black holes in AdS, $r_h \gg \ell$, one gets
\begin{equation}
b_c T \approx \left(\frac{d-1}{4\pi}\right)\frac{r_h}{\ell} \gg 1.
\label{b_c T}
\end{equation}
This implies that the geometric optics approximation is valid for quanta with energy comparable to the Hawking temperature, $\omega \sim T$, as there are many different values of the angular momentum that can freely fall into the black hole.

To put it another way, for $r_h \gg \ell$, $b_c T = r_c T_c \gg 1$, where $T_c = T V(r_c)^{-1/2}$ is the local temperature measured by a static observer at $r_c$, which is also the radius of the circular photon orbits.  The Hawking quanta at $r = r_c$ will have a typical wavelength $\sim 2\pi/T_c$, which is much less than the circumference $2\pi r_c$ of the sphere at the location of the circular photon orbits.  Therefore, one can view this sphere as emitting Hawking quanta that have wavelengths short in comparison with the size of the sphere.  Any such massless quanta emerging outward from this sphere will obey the geometrical optics approximation and propagate freely outward as Hawking radiation, with negligible probability of being backscattered at radii beyond this sphere, unlike the case for a `small' black hole as in asymptotically flat spacetime, where there is significant backscatter even outside the critical sphere (at $r = 3 GM$ for the 4-dimensional Schwarzschild metric), making numerical calculations of the absorption probabilities necessary for such black holes.

When the geometric optics approximation is valid for the bulk of the thermal Hawking radiation, as it is for `large' black holes in AdS with $r_h \gg \ell$, one can get the Hawking luminosity or power as the product of the Stefan-Boltzmann constant $\sigma_d$ in $d$ dimensions, the $d$ power of the Hawking temperature $T$, and the area $\Omega_{(d-2)}b_c^{d-2}$ of a sphere of the radius of the critical impact parameter.  (One can also get the same power by taking the local power emitted outward from the sphere at $r = r_h$ and multiplying by two factors of $V(r_c)^{1/2}$ to convert the local energy and rate to an energy and rate with respect to the Killing vector $\partial/\partial t$ normalized so that $g_{00}g_{rr} = -1$).

To get the Stefan-Boltzmann constant $\sigma_d$ without bothering to try to find it in the literature, one can first do a standard analysis to get the energy density of massless bosonic and fermionic field in a thermal state in a region of space of size and curvature lengths large compared with the thermal wavelengths as $a_d T^2$ with radiation constant
\begin{equation}
a_d = (d-1)\pi^{-\frac{d}{2}}\Gamma\left(\frac{d}{2}\right)\zeta(d)N,
\label{a}
\end{equation}
where $\zeta(d)$ is the Riemann zeta function, the sum of the inverse $d$ powers of the positive integers, and where $N$ is the number of bosonic modes (e.g., the number of polarizations for each species, summed over the species) plus a fraction $1-2^{1-d}$ of the number of fermionic modes.

Now the flux per unit area from a surface is $\sigma_d T^2$, where $\sigma_d/a_d$ is the ratio of the area of a unit $(d-2)$-dimensional disk to the area of a unit $(d-2)$-dimensional sphere, which is $\Omega_{(d-3}/[(d-2)\Omega_{(d-2)}]$.  This then gives the Stefan-Boltzmann constant in $d$ dimensions as
\begin{equation}
\sigma_d = \pi^{-\frac{d+1}{2}}\Gamma\left(\frac{d+1}{2}\right)\zeta(d)N.
\label{sigma}
\end{equation}

For one species of massless spin-1 bosons (e.g., photons) in $d$ dimensions, the number of polarizations is the number of independent transverse vectors, $N_1 = d-2$.  For one species of massless spin-2 bosons (e.g., gravitons) in $d$ dimensions, the number of polarizations is the number of independent transverse traceless symmetric tensors, $N_2 = d(d-3)/2$.  If one has one species of both photons and gravitons, one would have $N = N_1 + N_2 = (d^2 - d - 4)/2$.  In $d=4$ this gives $N=4$, two photon polarizations and two graviton polarizations.

Then the Hawking emission power is
\begin{equation}
-\frac{dM}{dt} = P = 4 \left(\frac{d-1}{4\pi}\right)^{d+1} \zeta(d) N \frac{1}{\ell^2} x^d y^d z^{-\frac{d-2}{2}} f,
\label{P}
\end{equation}
where
\begin{equation}
x \equiv \frac{r_h}{\ell},
\label{x}
\end{equation}
\begin{equation}
y \equiv 1 + \frac{d-3}{d-1}\frac{\ell^2}{r_h^2} = 1 + \frac{d-3}{d-1}\frac{1}{x^2},
\label{y}
\end{equation}
\begin{eqnarray}
z &\equiv& 1 + \frac{d-3}{d-1}\frac{\ell^2}{r_h^2}\left[\frac{d-1}{2}\left(\frac{r_h^2}{\ell^2}+1\right)\right]^{-\frac{2}{d-3}}\nonumber\\
 &=& 1 + \frac{d-3}{d-1}\frac{1}{x^2}\left[\frac{d-1}{2}(x^2+1)\right]^{-\frac{2}{d-3}},
\label{z}
\end{eqnarray}
and $f$ is the thermal-averaged cross section divided by the geometric optics value of $\Omega_{(d-3)} b_c^{d-2}/(d-2)$.  For $x \equiv r_h/\ell \gg 1$, the geometric optics approximation is good, so $y \approx z \approx f \approx 1$ and the Hawking emission power is proportional to $x^d\equiv (r_h/\ell)^d$.

Now when one evaluates $dM = TdS$ either from Eq.\ (\ref{S}) or from Eq.\ (\ref{M}) with $T = [(d-1)/(4\pi\ell)]xy$, one gets
\begin{equation}
-\frac{dM}{dt} = \frac{(d-1)(d-2)\Omega_{(d-2)}}{16\pi G}\,\ell^{d-3}\,x^d\,y
 \,\frac{d}{dt}\left(\frac{1}{x}\right).
\label{dM/dt}
\end{equation}
When one sets this equal to the formula given for the power in Eq.\ (\ref{P}), one gets
\begin{equation}
dt = \left(\frac{4\pi}{d-1}\right)^d \frac{(d-2)\Omega_{(d-2)}}{\zeta(d)Nf}\, \frac{\ell^{d-1}}{16G}\,y^{-(d-1)}\,z^\frac{d-2}{2}\,d\left(\frac{1}{x}\right).
\label{dt}
\end{equation}

For $x \equiv r_h/\ell \gg 1$ so that $y \approx z \approx f \approx 1$, a black hole that started at infinite initial size ($r_h \equiv \ell\,x = \infty$) and infinite initial mass ($M = (d-2)\Omega_{(d-2)}\ell^{d-3}(x^{d-1}+x^{d-3})/(16\pi G) = \infty$)
at $t=0$ would evolve in finite time $t$ down to a finite size,
\begin{equation}
x \equiv \frac{r_h}{\ell} \approx \frac{(d-2)\Omega_{(d-2)}}{16\zeta(d)N}\,
\left(\frac{4\pi}{d-1}\right)^d\,\frac{\ell^{d-1}}{G}\,t^{-1},
\label{xx}
\end{equation}
and to a finite mass,
\begin{equation}
M \approx \frac{\zeta(d)N}{\pi}\,
  \left[\frac{(d-2)\Omega_{(d-2)}}{16\zeta(d)N}\right]^d\,
  \left(\frac{4\pi}{d-1}\right)^{d(d-1)}\,
  \frac{\ell^{(d-2)(d+1)}}{G^d}\,t^{-(d-1)}.
\label{mass}
\end{equation}

Eqs.\ (\ref{x}) and (\ref{mass}) na\"{\i}vely suggest that it would take an infinite time $t$ for the black hole to evaporate to zero size and mass, but of course these equations are valid only for large black holes, $x \gg 1$.  One cannot integrate Eq.\ (\ref{dt}) into the regime of small $x$ without knowing how the factor $f$ (the ratio of the thermally-averaged cross section to the geometric optics value) behaves as a function of $x$ when $x$ is not large and the geometric optics approximation is no longer valid.  However, $f$ will have a positive minimum value, which I shall denote as $f_m$, since black holes do evaporate even for small $x$ (e.g., $x=0$ for asymptotically flat spacetime with no negative cosmological constant).  Presumably this lower limit occurs at $x=0$, since it is plausible that the geometric optics approximation gets better and better as $x$ increases, so that $f$ increases monotonically to approach unity for large $x$.  My old numerical calculations \cite{Page:1976df,Page:1976ki,Page:2004xp} for photons and gravitons in four-dimensional Schwarzschild spacetime ($x=0$) give the value $f_m \approx 0.13395$ for $d-4$, assuming that indeed the minimum is at $x=0$.

One can easily show that $z < y$ for all $x$, so that if one defines
\begin{equation}
u \equiv \sqrt{\frac{d-3}{d-1}}\,\frac{\ell}{r_h}
  \equiv \sqrt{\frac{d-3}{d-1}}\,\frac{1}{x},
\label{u}
\end{equation}
then replacing $z$ by the larger value $y = 1 + u^2$ in Eq.\ (\ref{dt}) and $f$ by the smaller value $f_m$ gives, for positive $du$ (e.g., for a black hole that is evaporating),
\begin{equation}
dt < \frac{(d-2)\Omega_{(d-2)}}{16\zeta(d)N f_m}\,
     \left(\frac{4\pi}{d-1}\right)^d\,\sqrt{\frac{d-1}{d-3}}\,
     \frac{du}{(1+u^2)^{d/2}}.
\label{dt-bound}
\end{equation}
One can then integrate this from $u=0$ (infinite black hole size) to $u=\infty$ (zero black hole size) to get a finite upper bound on the lifetime $\Delta t$ of an initially infinitely large and massive black hole evaporating in asymptotically anti-de Sitter spacetime with absorbing boundary conditions at infinity:
\begin{equation}
\Delta t < \sqrt{\frac{d-1}{d-3}}\,\left(\frac{4\pi}{d-1}\right)^d\,
   \frac{\pi^{d/2}}{\Gamma\left(\frac{d-2}{2}\right)}\,
   \frac{\ell^{d-1}}{8\zeta(d)N f_m G}.
\label{t-bound}
\end{equation}

If one now goes to four-dimensional spacetime and uses my numerical value $f_m \approx 0.13395$, one gets
\begin{equation}
\Delta t < \frac{80\sqrt{3}\pi^2}{9 f_m}\,\frac{\ell^3}{G}
   \approx 1134\,\frac{\ell^3}{G}.
\label{t-bound-4d}
\end{equation}
One can fairly easily do somewhat better in $d=4$ by using the exact expression for $z$ instead of replacing it by its upper bound $y$, though still replacing $f$ by $f_m$, giving
\begin{eqnarray}
\Delta t &<& \frac{320\sqrt{3}\pi\ell^3}{9 f_m G}
\int_0^\infty\,\frac{(1+3u^2)^2 + 4u^6}{(1+u^2)^3(1+3u^2)^2}\,du
\nonumber\\
&=& \frac{40\pi^2}{9 f_m}(8 - 3\sqrt{3})\,\frac{\ell^3}{G}
\approx 918\,\frac{\ell^3}{G},
\label{t--better-bound-4d}
\end{eqnarray}
which is $(8\sqrt{3} - 9)/6 \approx 0.8094$ times as large as the previous more crude limit.  The actual limit might be of the order of two or more times smaller, since $f$ is expected to be greater than $f_m$ for positive $x$ and approach $1 \approx 7.4655 f_m$ as $x$ gets large.  However, it would require extensive numerical calculations to evaluate $f(x)$ and do the integral to get a precise value for the upper bound on the total decay time even in just four-dimensional spacetime, which is beyond the scope of this present paper that just establishes the existence of a finite decay time for an initially infinitely large and infinitely massive black hole in asymptotically anti-de Sitter spacetime without giving a precise numerical value for this decay time.
 
In conclusion, arbitrarily large black holes in asymptotically anti-de Sitter spacetime do {\it not} take an arbitrarily long time to evaporate away; instead the total decay from infinite initial mass to zero final mass takes a finite time that has the form
\begin{equation}
\Delta t = C\,\frac{\ell^{d-1}}{G},
\label{t-bound-again}
\end{equation}
where for the spherically symmetric black holes considered in this paper, $C$ is a finite constant that depends on the spacetime dimension $d$ and on the field content of the theory.  The value of $C$ for various cases remains to be determined numerically, since it depends on how the thermally-averaged cross section of the black hole for the various fields departs from the geometric optics value for black holes which are not large in comparison with the AdS length scale $\ell$. 

This calculation was motivated by discussions with Pisin Chen, Yasusada Nambu, Yen Chin Ong, Misao Sasaki, Dong-han Yeom, and others at the 2015 May 18-29 Molecular-Type Workshop YITP-T-15-01 on the Black Hole Information Loss Paradox at the Yukawa Institute for Theoretical Physics at Kyoto University, Japan, and was partially completed there; I greatly appreciated the hospitality and stimulating interactions.  This research is also supported in part by the Natural Sciences and Engineering Council of Canada.

\end{document}